\documentclass[apjl]{emulateapj}

\def\msun{M$_\odot$}

\slugcomment{Submitted November 2005; Accepted February 2006}

\shorttitle{On Stellar Multiplicity}
\shortauthors{C.J. Lada}

\begin{document}


\title{Stellar Multiplicity and the IMF: Most Stars Are Single}

\author{Charles J. Lada\altaffilmark{1}}
\altaffiltext{1}{Harvard-Smithsonian Center for Astrophysics, 60 Garden Street, 
Cambridge, MA 02138, USA; clada@cfa.harvard.edu}

\begin{abstract}

In this short communication I compare recent findings suggesting a low binary
star fraction for late type stars with knowledge concerning the forms of the
stellar initial and present day mass functions
 for masses down to the hydrogen burning limit.  This comparison
indicates that most stellar systems formed in the Galaxy are likely single and
not binary as has been often asserted.  Indeed, in the current epoch two-thirds
of all main sequence stellar systems in the Galactic disk are composed of single
stars.  Some implications of this realization for understanding the star and
planet formation process are briefly mentioned.

\end{abstract}


\section{Introduction} \label{sec:introduction}

Ever since Mitchell (1767) pointed out that the observed frequency of visual
double stars was too high to be due to random chance, the study of binary stars
has occupied an important place in astrophysics.  William Herschel (1802)
discovered and cataloged hundreds of visual pairs and produced the first
observations of a rudimentary binary orbit.  In doing so he established that the
double stars were indeed physical pairs and that Newtonian physics operated
nicely in the distant sidereal universe.  By the beginning of the twentieth
century tens of thousands of binary stars were known and cataloged (e.g., Burnham
1906).  By the middle to late twentieth century the first systematic attempts to
establish the binary frequency of main sequence F and G stars suggested that a
very high fraction (70 - 80\%) of all such stellar systems consist of binary or
multiple stars (Heintz 1969; Abt \& Levy 1976; Abt 1983).  The most comprehensive
and complete study of the multiplicity of G stars was performed by Duquennoy \&
Mayor (1991) who argued that two-thirds of all such stellar systems are
multiple.

It has often been assumed but never clearly demonstrated that similar statistics
applied to stars of all spectral types.  This assumption has led to the commonly held
opinion that most all stars form in binary or multiple systems with the Sun (and its
system of planets) being atypical as a single star. 
But how robust is the assumption
that the binary statistics for G stars is representative of all stars?

Over the last decade two important developments have occurred in stellar research which
directly bear on this question.  First, the functional form of the stellar initial mass
function (IMF) has been better constrained by observations of both field stars (e.g.,
Kroupa, 2002) and young embedded clusters (e.g., Muench et al.  2002).  The IMF has been
found to peak broadly between 0.1 - 0.5 \msun, indicating that most stars formed in the
Galactic disk are M stars.  Second, surveys for binary stars have suggested that the binary
star frequency may be a function of spectral type (e.g., Fischer \& Marcy 1992).  In
particular, there have been a number of attempts to ascertain the binary frequency of M type
stars and even for L and T dwarfs, objectss near and below the hydrogen burning limit.  These
studies suggest that the binary frequency declines from the G star value, being only around
30\% for M stars (e.g., Leinert et al.  1997; Reid \& Gizis 1997; Delfosse et al.  2004;
Siegler et al.  2005) and as much as a factor of 2 lower for L and T dwarfs (e.g., Gizis et
al.  2003).  I argue in this communication that these two facts together suggest that most
stellar systems in the Galaxy consist of single rather than binary or multiple stars.


\section{The Single Star Fraction and Spectral Type}
\label{sec:observations}

In this section I use data compiled from the literature to examine the single
star fraction as a function of stellar spectral type, in particular for the range
spanning G to M stars.  I consider the single star fraction (SSF) to be the
fraction of stellar systems without a {\it stellar} companion, that is, primary
stars without a companion whose mass exceeds 0.08 \msun.  Figure 1 displays the
single star fraction as a function of spectral type for G and later type stars.
This plot suggests that the SSF is significantly greater for M stars than for G
stars.  Indeed the SSF for M stars appears to be at least 70\%.  It is difficult
to evaluate the significance of this difference at face value given that the
differing binary surveys suffer from differing biases and varying degrees of
incompleteness.  The systematic differences that can arise between the surveys
mostly derive from varying sensitivities to primary/secondary separations and 
mass ratios.  Below I attempt to evaluate the results from the surveys used to
construct Figure 1.

\begin{figure}
\centering
\includegraphics[scale=0.40]{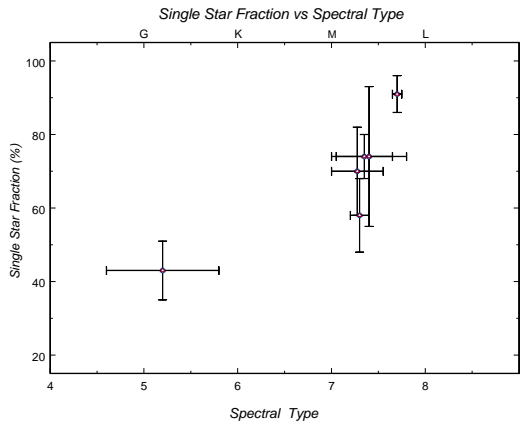}
\caption{The single star fraction vs spectral type. The single star
fraction increases significantly with spectral type reaching values
of $\sim$ 75\% for M stars, the most populous stars in the IMF and 
the field. Vertical error bars represent statistical uncertainties in
the SSF. The horizontal error bars indicate the approximate extent in
spectral type covered by the individual surveys and do not represent
an uncertainty in this coordinate. Data taken from 
Duquennoy \& Mayor (1991), Reid \& Gizis (1997), Fischer \& Marcy (1992), 
Delfosse et al. (2004), Leinert et al. 1997, and Siegler et al. (2005).
}
\end{figure}

In their seminal study, Duquennoy \& Mayor (1991) obtained a spectroscopic survey
of a distance-limited complete sample of F7-G9 stars in the Northern Hemisphere
and within 22 pc of the Sun.  They examined radial velocities obtained for these
stars over a 13 year period.  They combined their detections of spectroscopic
binaries with known visual binaries and common proper motion pairs to examine 164
primaries for evidence of multiplicity.  They derive multiplicity ratios of
57:38:4:1 for single:double:triple:quadruple systems, respectively.  They
considered all the various detection biases to estimate the incompleteness of
their study and concluded that there was a slight bias against detecting low mass
companions, this resulted in a 14\% upward correction to the multiplicity
fraction such that 57\% of systems were estimated to be multiple for a
primary/companion mass ratio, q $>$ 0.1.  They further extrapolated this
incompleteness correction to include substellar secondaries and estimated a
multiplicity fraction of 2/3 and a single star fraction of 1/3 for their sample.
However, in recent years sensitive and precise radial velocity surveys of 1330
single FGKM stars have indicated a paucity of substellar companions within 5 AU
of the primary stars (Marcy \& Butler 2000; Marcy et al.  2005).  In addition
coronographic imaging surveys have found a similar dearth of substellar
companions around GK and M stars over separations between 75 and 300 AU (McCarthy
\& Zuckerman 2004).  The existence of this so-called ``brown dwarf desert''
indicates that Duquennoy \& Mayor may have overestimated the multiplicity
fraction of G stars and the true value is likely 57\% or even somewhat smaller.
For the purposes of this paper I adopt 57\% as the multiplicity fraction of G
type stars and thus 43\% for the SSF.

The first extensive examination of the multiplicity of M stars was performed by Fischer \&
Marcy (1992) who studied radial velocity, speckle and visual binary data for a sample of
stars within 20 pc.  The full range of separations, $a$ $<$ 10$^4$ AU, was examined, similar
to the G star study.  These authors pointed out that M star surveys suffer less from the
effects of incompleteness than G star surveys because the M star sample is on the whole a
factor of 2 closer in distance and M star primaries are sufficiently faint to enable
detection of very faint companions more readily.  They derived a SSF of 58\% which is higher
than the G star value.


Reid \& Gizis (1997) determined the SSF for a volume complete sample of 79
M2-M4.5 primary stars within 8 pc of the Sun and derived a SSF of 70 $\pm$ 12\%
for this sample.  The range of binary separations they were able to probe was 0.1
- 10$^4$ AU.  A similar volume complete search for M dwarf binaries within 5 pc
of the Sun was performed by Leinert et al.  (1997) who reported a SSF of 74 $\pm$
19\%.  However, their sample of 29 stars is smaller than the Reid \& Gizis (1997)
and Fischer \& Marcy (1997) samples accounting for the larger uncertainty.  More
recently Delfosse et al.  (2004) presented statistics for a much larger sample of
100 M dwarfs which they estimated was 100\% complete for stellar mass companions
over the entire separation range and out to 9 pc from the Sun.  Delfosse et al.
(2004) derive a multiple star fraction of 26 $\pm$ 3 \% which corresponds to a
SSF of 74 $\pm$ 6\%.  This may represent the most accurate determination for the
M star SSF yet made.  I note here that even if one considers substellar
companions this estimate for the SSF will not likely alter significantly since as
mentioned earlier, surveys have revealed a dearth of substellar companions to G,
K {\it and} M stars (Marcy \& Butler 2000; McCarthy and Zuckerman 2004).

Surveys for multiplicity among very late M stars and even L and T dwarfs have
also been recently reported.  These studies typically explore more limited
separation ranges and somewhat smaller samples of stars.  The multiplicity
fractions they find are however all lower than that reported for the earlier type
M stars.  For example, Siegler et al.  (2005) examined a magnitude-limited survey
of 36 M6 - M 7.5 stars and derived a binary fraction of 9 $\pm$ 4\% corresponding
to a SSF of 91 $\pm$ 5\%.  However this sample is not volume limited and may be
incomplete.  Thus the inferred SSF is likely an upper limit.  Despite this
limitation Siegler et al.  were able to conclude that wide (a$>$20 AU) binaries
are very rare among these stars.  Although not considered for inclusion in Figure
1 because of the large fraction of brown dwarfs in their samples, surveys by
Gizis et al.  (2003) and Bouy et al.  (2003) find similarly small binary
fractions for ultra low mass objects.  For example, Gizis et al.  examined 82
nearby late M and L dwarfs and derived a (incompleteness corrected) binary
fraction of 15 $\pm$ 5\% (corresponding to a SSF of 85 $\pm$ 14\%) for 
separations, $a > $ 1.6 AU. Estimating the possible contribution of
companions at smaller separations they suggest a binary star fraction (BSF)
of 15 $\leq$ BSF $\leq$ 25 \%  corresponding to 75 $\leq$ SSF $\leq$ 85 \%
for these
objects near and just below the hydrogen burning limit.  Bouy et al. (2003)
examined the binary statistics for a sample of 134 late M and L field dwarfs and
estimated a binary fraction for a separation range of about 2 - 140 AU of only
10\% corresponding to a SSF of 90\% for these objects.  They also noted a dearth
of companions with wide (i.e., $a >$ 15 AU) separations.  Although these surveys
of very low mass and substellar objects suffer from some degree of incompleteness
it is quite unlikely that sensible corrections for such effects would decrease
the estimated single star fraction to a value similar to that of G stars or even
typical M stars.

The observations discussed above lead to the conclusion that the single 
star fraction is a function of spectral type and increases from about
43\% for G stars to $\sim$ 85\% for brown dwarfs. The most secure estimate
for M stars appears to be about 74\% based on the complete volume-limited
sample of Delfosse et al. (2004) for M stars with stellar companions.

\section{M Stars and the IMF} \label{sec:results}

The stellar IMF is one of the most fundamental
distribution functions in astrophysics.  A great deal of effort has been expended
in determining its form since the first attempt to measure its shape by Salpeter
(1954).  He found that the IMF is a power-law which decreases with stellar mass
for field stars with masses in the range between 1-10 \msun.  More recent
determinations of the IMF for field stars and young embedded clusters have
expanded the mass range covered by Salpeter.  These studies have found the IMF
to break from a single power-law shape near 0.5 \msun\ and to have a broad peak
between $\sim$ 0.1 - 0.5 \msun.  On either side of this peak the IMF falls off
rapidly (e.g., Miller \& Scalo 1979; Kroupa 2002; Muench et al. 2002; Chabrier
2003; Luhman et al. 2006). 

\begin{figure}
\centering
\includegraphics[scale=0.36, angle=90.0]{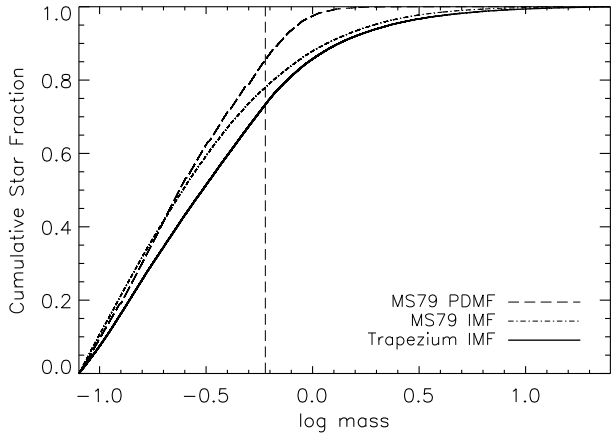}
\caption{The cumulative frequency distributions for all hydrogen burning
stars in two versions of the primary star IMF and the 
PDMF of main sequence field stars. The two IMFs correspond to the
Miller-Scalo field star IMF and the IMF derived for the young embedded
Trapezium cluster by Muench et al. (2002). The vertical line marks the location
of the M star boundary (Torres \& Ribas 2002). The fraction of M stars is high for all these
mass functions ranging between 73 and 84\%. The latter value representing
fraction of all main sequence field stars that are M stars currently 
residing in the Galactic disk. Based on data from Miller \& Scalo 1989 and
Muench et al. 2002.}
\end{figure} 

The broad peak of the IMF encompasses the M stars and indicates that these stars
are the most numerous objects created in the star formation process.  This is
illustrated in Figure 2 which shows the cumulative fraction of all stars above
the hydrogen burning limit given by the IMF.  Two different IMFs are plotted
which span the range of modern day determinations of this function.  One is the
log-normal field star IMF derived by Miller \& Scalo (1979) and the other
represents a determination of the IMF for the embedded Trapezium cluster in Orion
in which the IMF is characterized by a series of broken power-laws (Muench et al.
2002).  This latter IMF is very similar to that determined for the field by
Kroupa (2002) but is more sensitive to substellar masses (not plotted).  The
vertical dashed line shows the boundary for the M star population.  The fraction
of all stars {\it above the hydrogen burning limit (HBL)} that are M stars is
73\% for the Muench et al.  IMF and 78\% for the Miller-Scalo IMF.  (It is
important to note here that these two IMFs are essentially primary star IMFs,
that is, IMFs that do not include companion star masses.)  This analysis
indicates that roughly 3/4 of all stars formed are M stars.

The IMF represents the frequency distribution of stars at birth and differs from
the present day mass function (PDMF) which represents the frequency distribution
of all stars currently living within the Galactic disk.  Stellar evolution has
significantly depleted the high mass end of the PDMF relative to the IMF.
Therefore, the fraction M stars in the PDMF is somewhat higher than the fraction
in the IMF.  Indeed, for the PDMF derived by Miller \& Scalo (1979) we find from
Figure 2 that 84\% of all stars in the Galactic disk are M stars.

\vskip 0.2in

\section{The Total Single Star Fraction} \label{sec:SSF}


To estimate the total fraction of single stars, I assume that all stars earlier
than M are characterized by the single star fraction for G stars determined by
Duquennoy \& Mayor (1991), that is, $SSF_{<M} =$ 43\%.  The single star
fraction for M-type stars (i.e., $SSF_{M}$) is assumed to be that (74\%)
determined by Delfosse et al.  (2004) for a complete, volume limited sample.  The
total SSF is then simply given by:

\small
$$ {\rm SSF(total)} = SSF_{<M} \times ETF + SSF_{M} \times MTF $$
\normalsize

\noindent 

Here $MTF$ is the M-type fraction, that is, the fraction of all stars that are M-type stars
and $ETF = 1 - MTF$ is the early-type fraction, that is the fraction of all stars that have
spectral types earlier than M.  To determine the SSF for all stars produced at any one time
by the star formation process I adopt the Muench et al.  and Miller-Scalo IMFs,
specifically, MTF = 0.73 and 0.78, respectively.  The total SSF is found to be 66\% and 67\%
for these two IMFs, respectively.  Therefore, single stars must ultimately account for as
many as two-thirds of all stellar systems that formed at any one time in the Galaxy.
Similarly, if we consider the MTF (0.84) for the Miller-Scalo PDMF we find the total SSF to
be 69\%.  Thus, {\it two thirds of all (main sequence) primary stars currently residing in
the Galactic disk are single stars}.

\section{Discussion and Conclusions}
\label{sec:discussion}



The primary result of this paper is the recognition that most stellar systems in
the Galaxy consist of single rather than binary stars.  This fact has important
consequences for star and planet formation theory.  For example, contrary to the
current accepted paradigm that most, if not all, stars form in binary or
multiple systems (e.g., Larson 1972, 2001; Mathieu 1994), this result could
indicate that the theoretical frameworks developed to explain the formation of
single, sunlike stars (e.g., Shu, Adams \& Lizano 1987) have wide applicability.
Indeed, when appropriately modified for a cluster-forming environment (e.g.,
Myers 1998; Shu, Li \& Allen 2004), they may even describe most star forming
events in the Galaxy.  On the other hand, most stars could still initially form
in binary or multiple systems provided that most such systems promptly
disintegrate via dynamical interactions or decay in an early, perhaps even
protostellar, stage of evolution (e.g., Kroupa 1995; Sterzik \& Durisen 1998,
Reipurth 2000).  

The current paradigm that most, if not all stars, form in binaries was
strengthened by early multiplicity surveys of pre-main sequence (PMS) stars.  In
particular, surveys of the PMS population of the Taurus cloud indicated a binary
fraction that was twice that of field G stars (Ghez et al.  1993; Leinert et al.
1993; Reipurth \& Zinnecker 1993).  However, most field stars are now known to
have formed in embedded clusters, environments quite different than represented
by the Taurus PMS population (e.g., Lada \& Lada 2003).  Binary surveys of both
young embedded and Galactic clusters have revealed binary fractions
indistinguishable from that of the field (e.g., Petr et al.  1998; Duch\^ene,
Bouvier \& Simon 1999; Patience \& Duch\^ene 2001). The most simple and
straightforward hypothesis to explain these two facts and the finding of
a high SSF in this paper is that the most common outcome of the star formation
process is a single rather than multiple star.


Observations of dust emission and extinction of molecular cloud cores have found
that the shape of the primordial or dense core mass function is very similar to
that of the stellar IMF except that the core mass function is offset to higher
mass by a factor of 2-3 (e.g., Stanke et al.  2005, Alves, Lombardi \& Lada
2005).  These observations indicate that a 1-to-1 mapping of core mass to
stellar mass, modified by a more or less constant star formation efficiency of
30-50\%, is possible, if not likely.  This idea is consistent with single
star systems being most often produced once the cores undergo collapse.

The fact that stellar multiplicity is a function of stellar mass, however, may provide
important clues to the nature of the physical process of star formation.  For example,
Durisen, Sterzik \& Pickett (2001) have shown that if individual protostellar cores can
further fragment and produce small N clusters, the dynamical decay of these clusters into
binary and single stars can in certain circumstances produce a binary star fraction that
declines with decreasing primary mass, similar to what is observed.  However, to be
consistent with the SSF derived here and to simultaneously produce reasonable binary
component separations, such models would require N $\geq$ 5, within a region $\sim$ 300
AU in size (Sterzik \& Durisen 1998).  This would correspond to a stellar surface density
($\sim$ 7.5 $\times$ $10^5$ stars pc$^{-2}$) about two orders of magnitude higher than the
peak density (7.2 $\times$ 10$^3$ stars pc$^{-2}$) measured for the rich Trapezium cluster
(Lada et al.  2004).  Such ultra-dense protostellar groups have not yet been identified, but
could be revealed with high resolution infrared imaging surveys of deeply embedded
candidates.  A related possibility, proposed by Kroupa (1995) and collaborators, posits that
all stars are formed in binaries in modestly dense embedded clusters.  Dynamical
interactions between these systems can disrupt some binaries and modify the separations of
others.  These models can produce the observed dependance of binary frequency with mass, but
at the expense of a SSF (50\%) that is too low to be consistent with that derived here.
These models could be made consistent with the high Galactic  SSF by assuming more compact
configurations for the birth clusters, however it is unclear whether 
the required higher cluster densities would remain consistent with observed values.

Another possibility is that binary star formation is related to the initial
angular momentum content of the primordial cores.  In this case the
initial angular momentum of a protostellar core would be expected to be a
function of core mass, with low mass cores being endowed with considerably less
angular momentum than high mass cores.  A systematic molecular-line survey of
cores of varying mass within a molecular cloud could test this idea.  A related
possibility is that turbulence may play a role in the propensity for a core to
fragment.  For example, Shu, Li \& Allen (2004) posit that the break in the
stellar IMF at 0.5 \msun\ is a result of the transition from turbulent to
thermal support of the envelopes of dense pre-collapse cloud cores.  The more
massive the core, the more turbulence is required to insure its support.
Ammonia observations of dense cores in fact do suggest that massive cores are
more turbulent than low mass cores (Jijina, Myers \& Adams 1999).  Perhaps
increased cloud turbulence in the more massive dense cores can also promote, in
some fashion, more efficient core fragmentation and a higher incidence of binary
star formation.  In this context it would be interesting to know if the trend of
increasing stellar multiplicity with stellar mass continues to the more massive
A, B and O stars, as has been suggested in some studies (e.g., Preibisch,
Weigelt, \& Zinnecker 2001, Shatsky \& Tokovinin 2002).

Finally I note that the large fraction of single star systems in the field is
consistent with the idea that most stars could harbor planetary systems
unperturbed by binary companions and thus extra-solar planetary systems
that are characterized by architectures and stabilities similar
to that of the solar system could be quite common around M stars, provided 
planetary systems can form around M stars in the first place.

\vskip -0.1in

\acknowledgments

I am indebted to August Muench for constructing the cumulative IMFs presented in
Figure 2 and many useful discussions.  I thank David Latham and Bo Reipurth for
their careful reading of the paper and detailed suggestions and Kevin Luhman,
Geoff Marcy, Frank Shu and Pavel Kroupa for useful comments which
improved the paper.

\end{document}